\documentclass[%
 reprint,
 amsmath,amssymb,
 aps,
]{revtex4-2}

\usepackage{graphicx}
\usepackage{dcolumn}
\usepackage{bm}

\begin{document}

\title{A Phase Field Crystal Method for Multilayer Graphene Structure}
\author{Kai Liu}
\affiliation{Center for Mathematical Sciences, Huazhong University of Science and Technology, Wuhan 430074, China}

 \date{\today}

\begin{abstract}
Bilayer graphene has been a subject of intense study in recent years. We extend a structural
phase field crystal method to include an external potential from adjacent layer(s), which is generated by the corresponding phase field and changes over time.
Moreover, multiple layers can be added into the structure. Using the thickness of the boundaries
between different stacking variants of the bilayer structure as the key parameter, we quantify the
strength of the adjacent layer potential by comparing with atomistic simulation results. We then test
the multiple graphene structures, including bilayers, triple layers, up to 6 layers. We find that besides the
initial conditions, the way of new layers added into the structure will also affect the layout of the atomic configuration. We believe tour results can help understanding the mechanism of graphene structure consists of more than one layer.

\end{abstract}

\maketitle

\section{Introduction\label{s1}}

Graphene, a single layer of carbon atoms tightly bound in a hexagonal honeycomb lattice, is one of the most exciting new two-dimensional materials discovered. Bilayer graphene has attracted a great deal of attention because it can exist with a variety of stacking arrangement with intriguing electronic properties \cite{Oostinga2008naturematerials,butz2014dislocations,ohta2006controlling,yankowitz2019tuning}.
In 2018, Yuan Cao et al. found that in twist angles of about 1.1$^\circ$ the electronic band structure of a twisted bilayer graphene exhibits unconventional superconductivity \cite{cao2018unconventional}, which draws a great amount of attention onto the bilayer graphene.

Computational modeling can serve as a route for theoretical understanding of the difficult-to-measure properties of graphene. On the continuum scale, the phase field crystal (PFC) modeling approach describes the dynamics of phase transformation through an atomically varying order parameter field that is loosely connected to the atomic density field.
The original PFC model was predominately used for the
study of 2D triangular and three-dimensional (3D) crystal symmetries \cite{elder2004modeling, elder2002modeling}. It is a
promising and widely used approach for modeling many microstructure phenomena. Recently, PFC has been used to study how anisotropic diffusion of carbon on a surface can yield the formation of the dendritic graphene structure \cite{meca2013epitaxial}. By including a rotationally invariant three-point correlation function for the excess free energy, a structural PFC model (XPFC) was set up to address both the atomically varying defect and microstructures of graphene and its nucleation and diffusional growth kinetics from a disordered state on a surface \cite{Seymour2016Structural,Hirvonen2016}.

In this paper, we build a new XPFC model for multilayer graphene by extending the XPFC method in \cite{Seymour2016Structural, Kai}.
In order to model the effect of one graphene layer onto another in an adjacent structure, we introduce a local interaction between the order parameter density and an external potential. The external layer potential we use in this paper is similar to the first-principles calculations of the generalized stacking fault energy (GSFE) in bilayer graphene from reference \cite{Zhou2015Van, Kai}, however we use a variant of
the phase filed density instead. This potential is based on the phase field of the corresponding layer and changes over time, different from our previous work that the bottom layer was fixed (as if on a deposition substrate)\cite{Kai}.  Moreover, multiple layers structure can be constructed easily.

In the numerical simulations, we first use a case of a long narrow ribbon domain, which consists of 4 parts: continuous AB and BA region each of nearly 50\% of the entire domain and two narrow transition
between them, to calibrate the contacted-layer potential \cite{Kai}. Here the so-called AB and BA stacking has one of the first layer's sublattice atoms (A or B) directly on top of its opposite sublattice atom (B or A) in the second layer, or collectively called Bernal stacking \cite{Alden2013Strain}. By comparing with atomistic simulations and previous work [11], we quantify the
strength of the external layer potential.

We then simulate the multilayer  structures. We test bilayers, trilayers, and multiple-layers, and so on. We test different initial conditions: well structured, i.e., without 5 or 7 rings and the hexagons are neatly arranged along the same direction, and randomly generated, i.e. a constant value with small perturbation of Gaussian noise. We also test various ways of construction: one layer after another, multiple layers after the bottom layer, and all layers at the same time, etc. It turns out that once a base layer is formed, layers on the top will be affected and AB (or BA) stacking is more likely to be formed. Moreover, the defect grain boundaries will emerge at similar locations. 
  
The paper is organized as follows. In Section 2, we introduce the math modeling and the numerical method. In Section 3, we test various cases of multilayer structures. And finally in section 4 we end with a discussion of this work and possible future applications.

\section{Modeling and Method\label{s2}}

We add an adjacent layer potential into the XPFC model \cite{Seymour2016Structural,Zhou2015Van}.
Let $\rho_i$, $i= 1,2,...,n$ describe the spatial phase density of carbon atoms for $n$ layers of graphene, repsectively. A dimensionless density field is then defined as $\psi_i=(\rho_i-\bar\rho_i)/\bar\rho_i$, where $\bar \rho_i$ is the reference density of a disordered phase around which a functional expansion of the free energy is carried out. The free energy of a crystallizing system reads as
\begin{equation}
F_{\text{total},i}=F_{\text{id}}(\psi_i)+F_{\text{ex,2}}(\psi_i)+F_{\text{ex,3}}(\psi_i)+\sum _{j\in \mathcal{K}_i}F_{j,i}(\psi_j,\psi_i)
\end{equation}
where  $F_{\rm id}$ is the ideal free energy, $F_{\rm ex,2}$ the two-point interactions, $F_{\rm ex,3}$ the three-point correlations \cite{Seymour2016Structural}, and $F_{j,i}(\psi_j,\psi_i)$ the adjacent layer potential. For $i=2,...,n-1$, $\mathcal{K}_i=\{i-1,i+1\}$, $\mathcal{K}_1=\{2\}$, and $\mathcal{K}_n=\{n-1\}$. $F_{\rm id}$ is given by
\begin{equation}
F_{\text{id}}  = \int {d{\bf{x}}\left\{ {{{\psi _i^2 } \over 2} - \eta {{\psi_i ^3 } \over 6} + \chi {{\psi_i ^4 } \over {12}}} \right\}} ,
\end{equation}
where $\eta$ and $\chi$ are dimensionless parameters and we simply set $\eta=\chi=1$.
 The two-point correlation is based on hard-sphere-like interactions and it is governed by \cite{Seymour2016Structural}
\begin{equation}
F_{\text{ex,2}}  =  - \frac{1}{2}\int {\psi_i({\bf{x}})\int {C_2 ({\bf{x}} - {\bf{x'}})\psi_i({\bf{x'}})d{\bf{x'}}d{\bf{x}}} }.
\end{equation}
Here $C_2$  is  the two-point correlation function defined as  \cite{Seymour2016Structural}
\begin{equation}
C_2 ({\bf{x}}) =  - \frac{R}{{\pi r_0^2 }}\text{\bf circ}\left( {\frac{r}{{r_0 }}} \right),
\end{equation}
where $r_0$ sets the cutoff for the repulsive term, $R$ sets the magnitude of the repulsion, and
\begin{equation}
\text{\bf  circ}(r) = \left\{ \begin{array}{l}
 1,\quad r \le 1, \\
 0,\quad r > 1. \\
 \end{array} \right.
\end{equation}
The three-point density correlation is rotationally invariant and robust enough to capture all crystal structures described through a single bond angle \cite{Seymour2016Structural}. It is governed by
\begin{align}
&F_{\rm ex,3}  = \nonumber\\
&  - \frac{1}{3}\int {\psi_i({\bf{x}})\int {C_3 ({\bf{x}} - {\bf{x'}},{\bf{x}} - {\bf{x''}})\psi_i({\bf{x'}})\psi_i({\bf{x''}})d{\bf{x'}}d{\bf{x''}}d{\bf{x}}} } .
\end{align}
Here the three point correlation function $C_3$ is defined by
\begin{equation}
C_3 ({\bf{x}} - {\bf{x'}},{\bf{x}} - {\bf{x''}}) = \sum\limits_i {C_s^{(k)} ({\bf{x}} - {\bf{x'}})C_s^{(k)} ({\bf{x}} - {\bf{x''}})},
\end{equation}
where $C_s^{(k)}$ in polar coordinate reads as  \cite{Seymour2016Structural}
\begin{align}
 C_s^{(1)} (r,\theta ) &= C_r (r)\cos (m\theta ), \\
 C_s^{(2)} (r,\theta ) &= C_r (r)\sin (m\theta ), \\
 C_r (r) &= \frac{X}{{2\pi a_0 }}\delta (r - a_0 ).
\end{align}
Here $X$ is a parameter defining the strength of the interaction, $a_0$ corresponds to the lattice spacing given by $r_0/a_0=1.22604$, and $m=3$ defines bond order of the crystal phase. For the graphene system, $R = 6$ and $X^{-1} = 0.4$ \cite{Seymour2016Structural}.

 
The adjacent layer potential reads as
\begin{equation}
 F_{j,i}  = \frac{1}{\lambda} \int {d{\bf{x}}\left[ \bar\psi_j({\bf{x}}) \psi_{i}({\bf{x}}) \right]},
\end{equation}
where $j\in \mathcal{K}_j $, $ \bar\psi_j({\bf{x}})$ is the corresponding potential field, and $\lambda$ parameterize the strength of the energy.\\
\begin{figure}[htb]
\begin{center}
\includegraphics[trim=5 5 5 5, width=3.2 in]{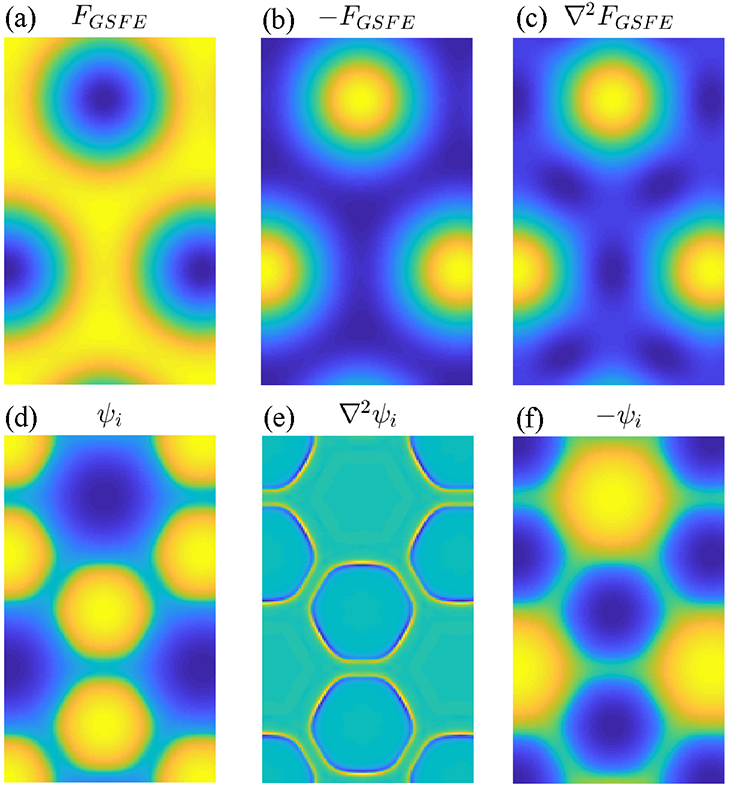}
\end{center}
\caption{(a) $F_{GSFE}$ from \cite{Zhou2015Van}, (b) $-F_{GSFE}$, (c) Non-dimensionalized $\nabla^2F_{GSFE}$. (d)   $\psi_i$. (e) Non-dimensionalized $\nabla^2\psi_i$, (f)    $-\psi_i$.
\label{PF}}
 \end{figure}\\
We then define $ \bar\psi_j({\bf{x}})$  as  
\begin{equation}
\nabla^2\bar\psi_i({\bf x})=D_0-D_1\psi_i({\bf x}),
\label{npsi}
\end{equation}
where  
\begin{equation}
D_0=\rm{mean}_{{\bf x}}\left(\nabla^2\bar\psi_i ({\bf x}) \right),
 \end{equation}
as shown in  Fig. \ref{PF}(f).   $\nabla^2 \psi_i$ is not good because the maximums and minimums of the $\nabla^2 \psi_i$ field are not at the locations of geometric center of the carbon rings or the center of carbon atoms, as shown in Fig. \ref{PF} (e).

Here we have compared different definition of $\nabla^2\bar \psi_j$, e.g. one that is mimic to $-F_{GSFE}$ in \cite{Zhou2015Van}, as shown by  Fig. \ref{PF} (b), which is better than $\nabla^2F_{GSFE}$ (used in \cite{Kai}), as shown in Fig. \ref{PF} (c). On the other hand,  the difference between $-F_{GSFE}$ and $-\psi$ are small and Eq. (\ref{npsi}) is more straightforward and stable in the application (see more details in the supplementary materials).

Finally, the evolution of the conserved density $\psi_i$ is governed by
\begin{equation}
\frac{{\partial \psi_i}}{{\partial t}} = M_{\psi_i} \nabla ^2  \left( \frac{{\delta F_{\text{total},i}}}{{\delta \psi_i}} \right),
\label{eom}
\end{equation}
where $M_{\psi_i}$ is an effective mobility that scales of the diffusional dynamics of $\psi_i$ and we set $M_{\psi_i}=1$ for convenience, if not specified.

We use a discrete Fourier transform (DFT) method to solve Eq. (\ref{eom}) \cite{Seymour2016Structural,Kai}. Period boundary condition is used for each layer, and a semi-implicit method is used to evolve Eq. (\ref{eom}) for numerical stability and computational efficiency.  The simulations are numerically expensive that it usually takes 3-7 days to reach a steady solidification. In order to solve the system in a large domain efficiently, we use a CUDA C/C++ , which runs about 2 orders of magnitude faster than the normal MATLAB CPU version.

\section{Results\label{s3}}


%
%
%

\subsection{Quantify $\lambda$}

Following \cite{Kai}, we use a long narrow ribbon of bilayer graphene to calibrate the parameter $\lambda$.
 \begin{figure}[htb]
 \begin{center}
  \includegraphics[  width=3.3in]{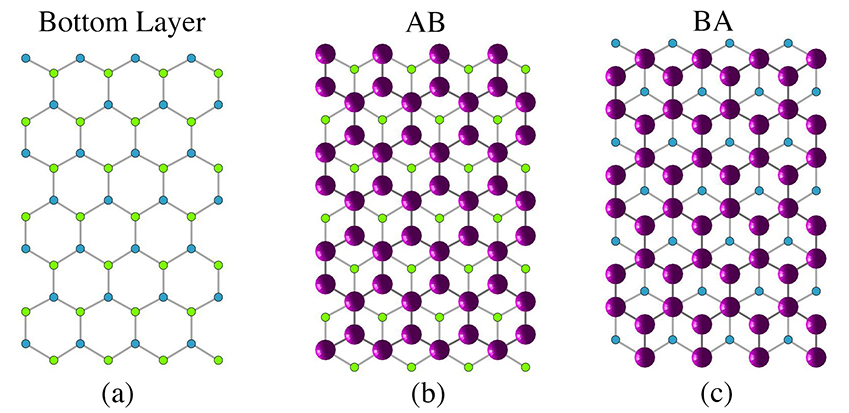}
\end{center}
\caption{(a) Bottom layer atoms are divided into 2 groups, one is denoted by lighter green and the other by darker green. (b) Position of atoms for AB stacking order.  (c) Position of atoms for BA stacking order.
\label{twopatterns}}
 \end{figure}
 
We first generate the AB and BA stacking phase field respectively, i.e. $\psi_1({\bf x})$, $\psi_{2,AB}({\bf x})$ and $\psi_{2,BA}({\bf x})$.  
Using well structured initial condition $\psi_1({\bf x},t_0)$ where $t_0=0$, i.e. ensuring the periodicity for each side of the rectangular domain (for example $L_1:L_2=2\times \sqrt{3}$) and all the hexagons are along the same direction, we generate the bottom phase field $\psi_1({\bf x})$ which agrees with \cite{Seymour2016Structural}. Next we $ \psi_1({\bf x},T)$ to form the adjacent layer potential. $\psi_2({\bf x},t_0)$ is a constant value (0.3) with small perturbation of Gaussian noise. Both AB and BA stacking are found, as shown in Fig. \ref{twopatterns}.  The long bilayer ribbon is then constructed by $\psi_1({\bf x})$ and $\psi_{2,AB}({\bf x})$ and $\psi_{2,BA}({\bf x})$.

The long  bilayer ribbon is set up as follows. The bottom layer $L1$ is formed by multiple small patch of $\psi_1({\bf x})$ $(L_1:L_2=2:\sqrt{3})$ , e.g. 32 patches, connected side to side.  $L2$ is then a static setup where there are 4 parallel stripes X-Y-Z-W.  X  represents the stacking order AB, e.g. $15\frac{1}{2}$ small patches of $\psi_{2,AB}({\bf x})$,  Z represents the stacking order BA, e.g. $15\frac{1}{2}$ patches of $\psi_{2,BA}({\bf x})$, and Y, W are disordered, i.e. constant value (0.3) with small perturbation.  The X and Z regions grow as the dynamics start and two interfaces will be created between them. 

\begin{figure}[ht]
 \begin{center}
\includegraphics[   width=3.3in]{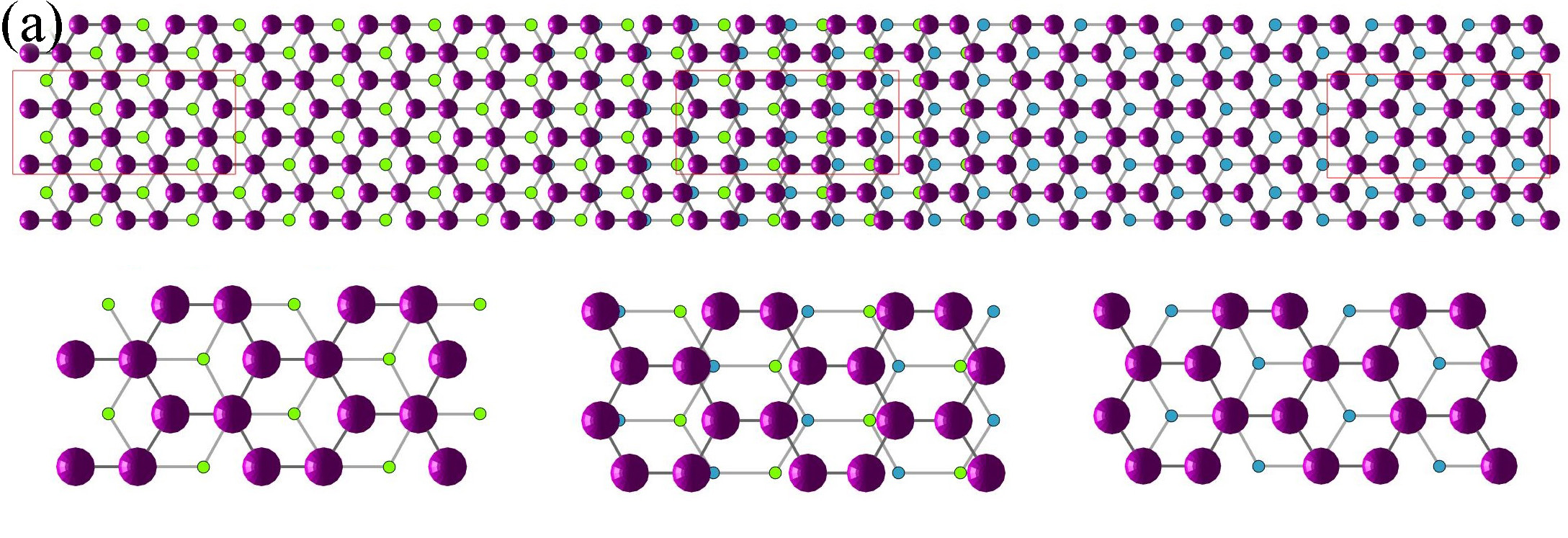}
\includegraphics[   width=3.3in]{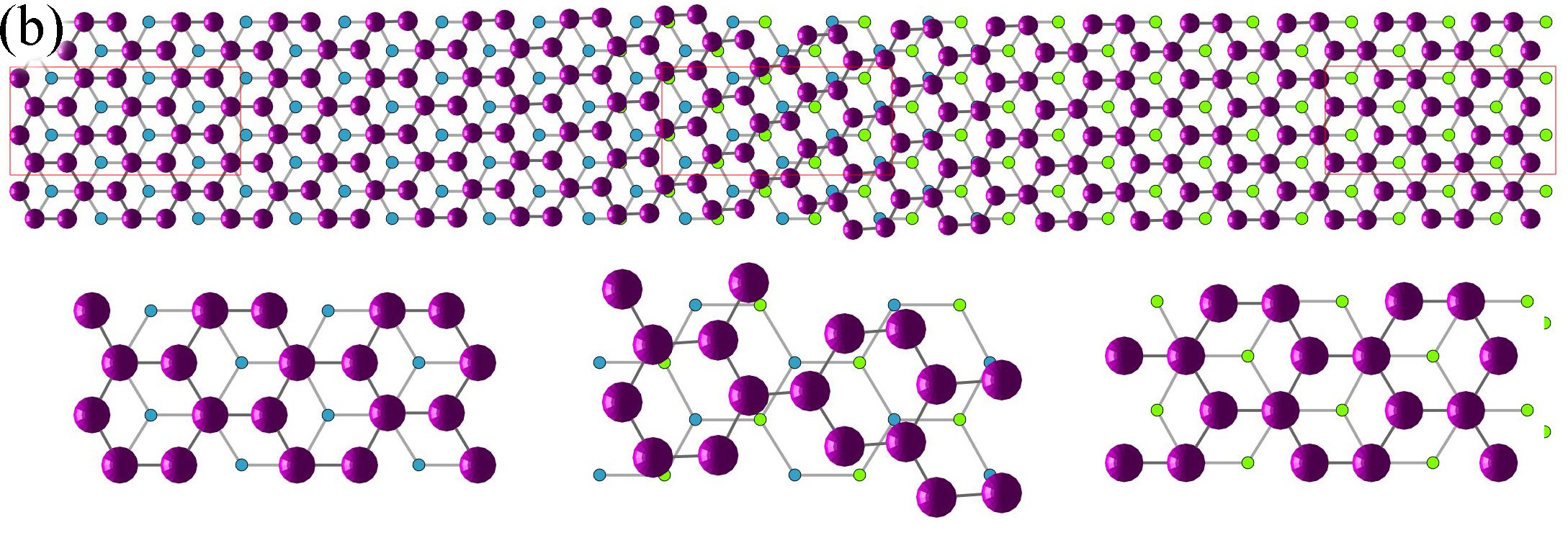}
\end{center}
\caption{(a) The AB to BA transition on the left region, (b) the BA to AB transition on the right region.
\label{riboexample}}
\end{figure}

We find four transition types, depending on the angle between the transition region and the shifting direction: 0$^\circ$, 30$^\circ$, 60$^\circ$, and 90$^\circ$. For example, the transition region is vertical in Fig. \ref{riboexample} (a) and (b),  and in Fig. \ref{riboexample} (a)  the angle is $90^\circ$ (the atoms shift horizontally) while in Fig. \ref{riboexample} (b) the angle  is $30^\circ$ (or $-30^\circ$). By comparing with the atomistic simulation results, we quantify the strength of the bottom layer potential by the width of the transition region for each type (angle).

\begin{figure}[ht]
\begin{center}
\includegraphics[width=3.5 in]{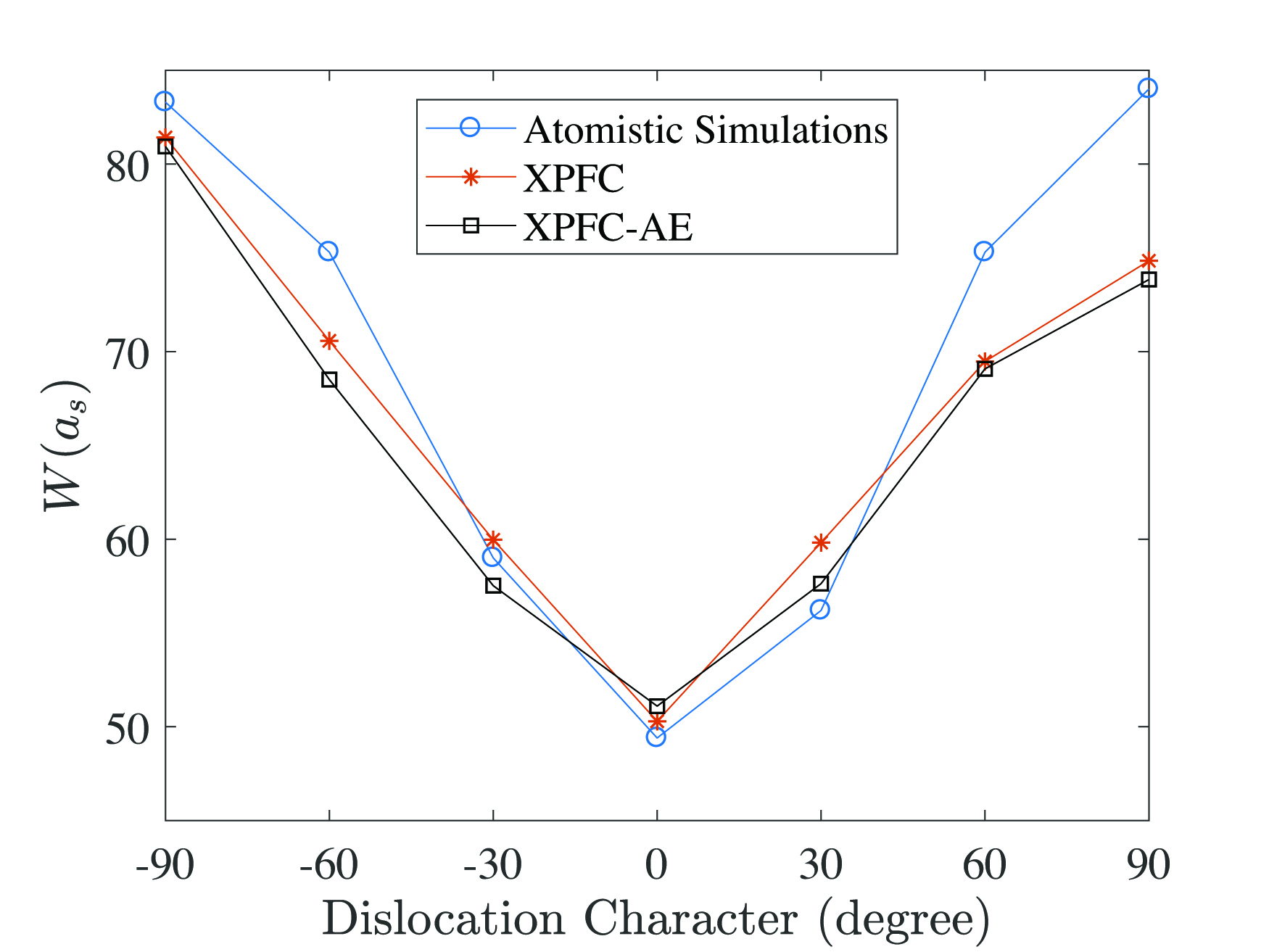} 
\end{center}
\caption{The relation between the dislocation direction and $W$. Here the angle between the direction of the transition region and the direction along which the atoms shift is used as the dislocation character. 
\label{Np}
}
 \end{figure}

We then compute the thickness of the transition region. A nondimensionalized parameter $d_{c_i}$,  the nondimensionalized  $x$-$y$ plane distance between the center of one group of atoms on the bottom layer and the nearest atom centers on the top layer at $x_i$, is defined to measure the distortion between substrate potential  $\psi_1({\bf x})$ and the graphene field $\psi_2({\bf x})$. For example, the distance in  $x$-$y$ plane between the center of the lighter green circles and the center of pink balls in Fig. \ref{riboexample}. Thus, $d_L=1$ for AB pattern and $d_L=0$ for BA pattern. Following \cite{Kai} the data is fitted by the function
 \begin{equation}
 d_L=\frac{2}{\pi}\arctan\left(\exp(\frac{\pi L}{W})\right) ,
 \label{fitlaw}
 \end{equation}
 where $W$ is a fitting parameter stands for the thickness of the transition region,  $L$ parameterizes the long side of the ribbon. Note that Eq.~(\ref{fitlaw}) gives an excellent fit to the displacements for both the atomistic and XPFC simulations \cite{Kai}.

Comparing with previous work \cite{Kai}, we've made two improvements. First we use a quadratic interpolation to calculate the exact position of the atoms, instead of using a specific grid point $(i, j)$. The  interpolation is performed along $x$ and $y$ direction separately as follows: suppose $\psi_{m,n}$ are the $k_{th}$ local maximums,  the maximum value among the $\{\psi_{i,j} | i\in \{m-k, m-k+1,...,m,..., m+k-1, m+k\}, j\in \{n-k, n-k+1,...,n,...,n+k-1,n+k\} \}$ (periodic conditions are implemented for points at the domain boundary). Here, $k$ is related to the size of the atoms.
 
Use the values of $\psi_{m,n}$,  $\psi_{m\pm1,n}$, and $\psi_{m,n\pm1}$, we perform a quadratic fitting. The position of the atom reads 
\begin{eqnarray}
\label{xposition}
X_{k}&=&m+ \frac{(\psi_{m,n+1}-\psi_{m,n-1})}{2(\psi_{m,n+1}+ \psi_{m,n-1}-2\psi_{m,n})},\\
Y_{k}&=&n+ \frac{(\psi_{m+1,n}-\psi_{m-1,n})}{2(\psi_{m+1,n}+ \psi_{m-1,n}-2\psi_{m,n})}.\label{yposition}
\end{eqnarray}
By this procedure, small change of the atoms' position can be computed accurately even if there is only slight change of phase field. Moreover,  a more accurate and efficient method is adopted to determine $W(t\rightarrow\infty)$. 

As discussed in \cite{Kai}, it takes a long time to reach a steady state for the long ribbon. On the other hand, $W(t)$ has an exponential tail behavior once $W(t)$ is close to the $W(\infty)$. We use 
\begin{equation}
W=W_0+b\exp(-ct)
\end{equation}
 to fit $W(t)$. Once $|W_0-W(T)|/W_0<0.001$,  $W_0$ is accepted.  
  
Use a bisection method, we find that $\lambda=15000$ is a proper value for the adjacent layer potential, considering all the three cases of $W_0$ and $W_{\pm90}$, as shown in Fig. \ref{Np}.  The values of 30 degrees and 60 degrees are not quite accurate, since the direction of the transition region might not be exactly perpendicular to the long side of the ribbon. 

\subsection{Multilayer Graphene Structure}

\subsubsection{Bilayer Structure}

\begin{figure}[ht]
\begin{center}
\includegraphics[trim=10 10 10 10, width=3.3 in]{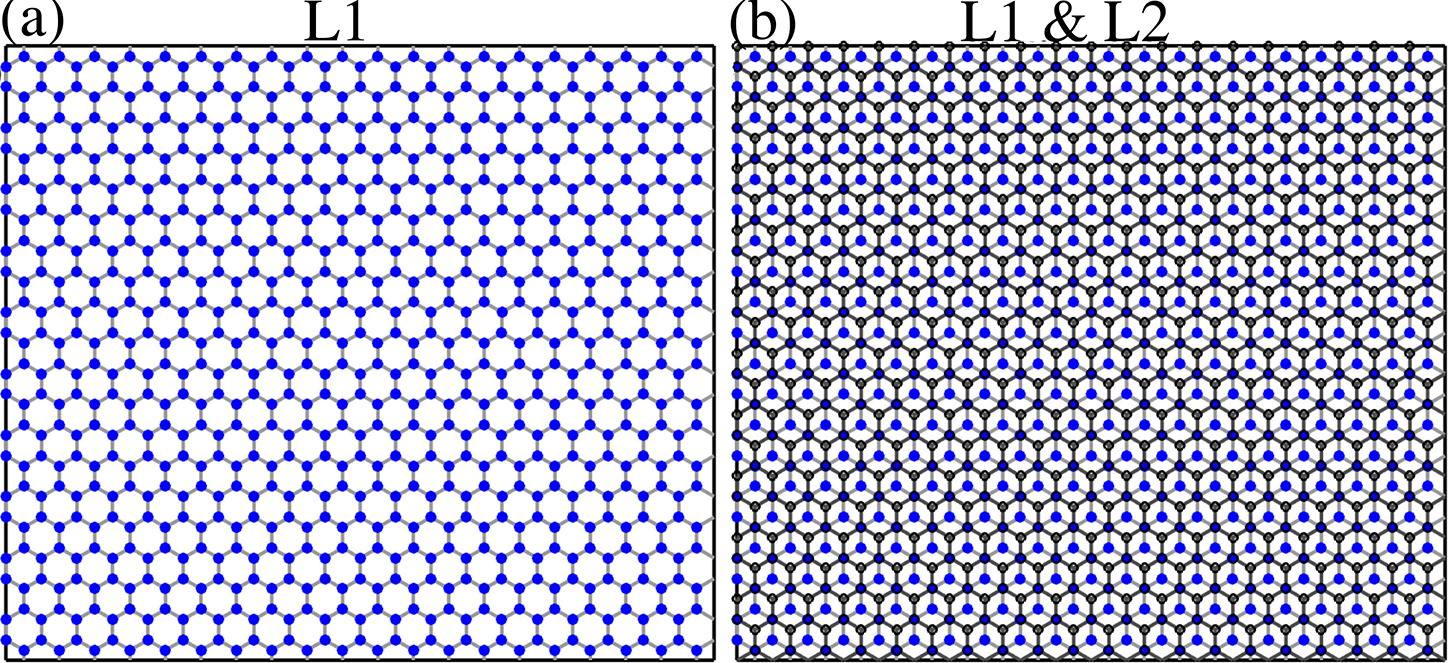}
\end{center}
\caption{ (a) The layout of the atomic configuration for the first (bottom) layer, (b)the two layers together (top on the bottom) where the blue dots are the atoms on the first layer and the black dots are the atoms on the top layer.
\label{twostructure}}
\end{figure}

Next we simulate the mutilayer graphene structures. We first test the bilayers. The first case is that the bottom layer ($L1$) is perfectly structured before the top layer ($L2$) is added into the system, as shown in Fig. \ref{twostructure} (a). As the solidification of $L2$ starts, the interaction between them is turned on, too. The $L2$ layer ends up with a perfect AB/BA stacking layout with $L1$, as shown in Fig. \ref{twostructure} (b). It shows that the bottom layer could determine the layout of upper layer, especially when it's well structured.
 
\begin{figure}[ht]
 \begin{center}
\includegraphics[trim=10 10 10 10, width=3.3in]{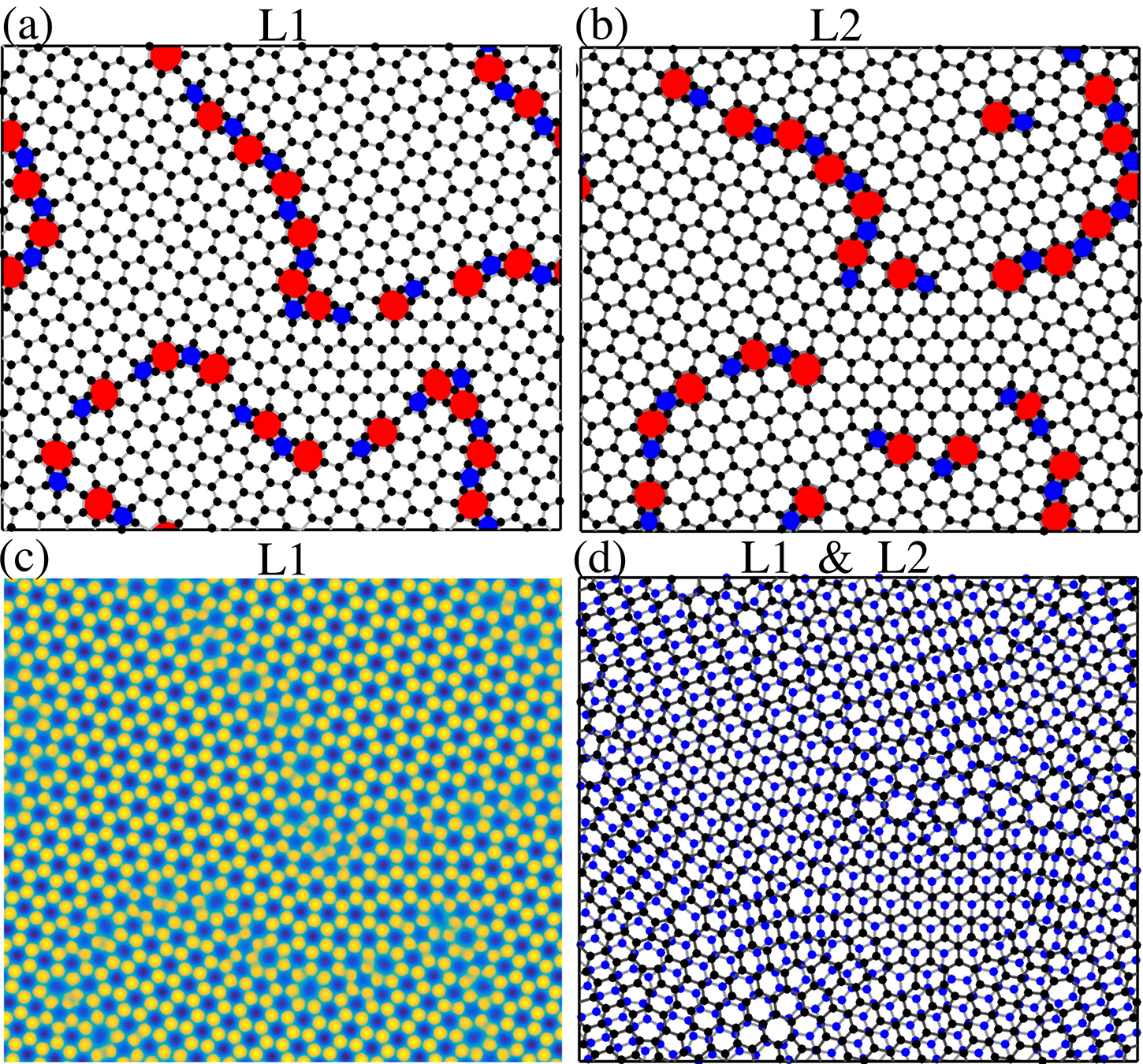}
\end{center}
\caption{ (a) the layout of the atomic configuration for $L1$, (b) layout of $L2$, (c) the phase field of  $L1$, and (d)  $L1$ \& $L2$ stacking, where the blue dots are the atoms on $L1$ and the black dots are the atoms on $L2$.
\label{twooneone}}
\end{figure}
\begin{figure}[htb]
 \begin{center}
\includegraphics[trim=10 10 10 10, width=3.3in]{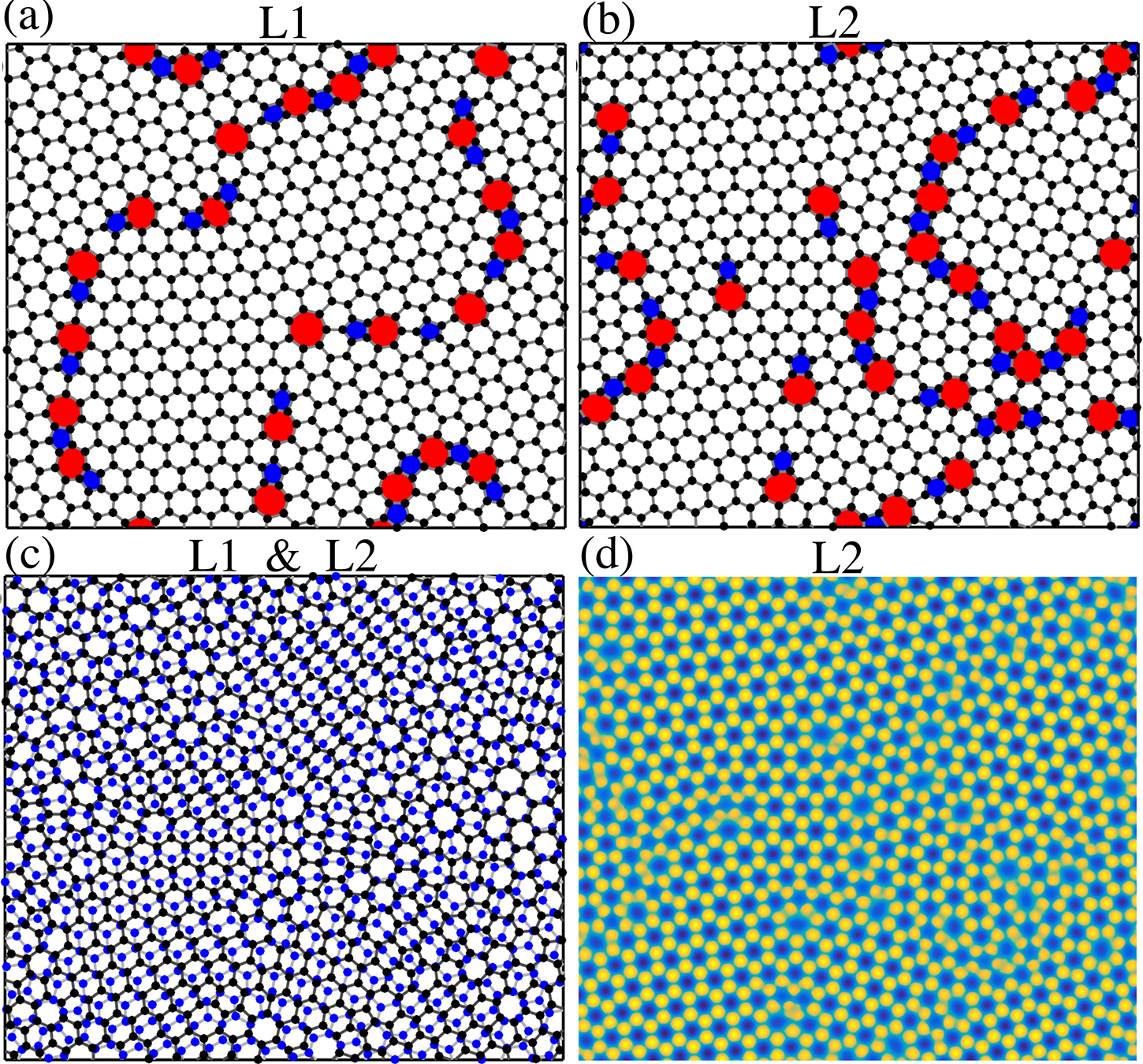}
\end{center}
\caption{ (a) the layout of the atomic configuration for $L1$, (b)  layout of  $L2$, (c)  $L1$ \& $L2$ stacking, where the blue dots are the atoms on $L1$ and the black dots are the atoms on $L2$, and (d) the phase field of $L2$.
\label{twosame}
}
\end{figure}

The second case is similar to the first one, except that the bottom layer $L1$ is solidified from a random initial condition, as show in Fig. \ref{twooneone} (a). The whole domain can be divided into two regions briefly, with 5/7 carbon rings along the grain boundary.  
Then the top layer begins to form with adjacent layer potential between $L1$ and $L2$. As shown in Fig. \ref{twooneone} (b) and $C$, the $L2$ is also highly affected by the bottom layer $L1$, that $L2$ can be briefly divided into two regions, too. The shape and location of the grain boundary, i.e. the 5/7 carbon rings, is close to those on $L1$. More than half of the entire domain are (close to) AB/BA stacking, and exceptions locate around the grain boundary, as shown in Fig. \ref{twooneone} (d) (see more details  in the supplementary materials).

The last case is that the solidification of the two layers start at the same time with different initial random conditions. It is an interesting case helps understanding of the limit of the effect of the adjacent layer potential. The results are shown in Fig. \ref{twosame}, that the locations of the 5/7 rings of the two layers are not strongly related. Also, there are only small regions close to the AB/BA stacking order, as shown in Fig. \ref{twosame} (c). The layout of the atomic configuration can be divided into several patches by the grain boundaries composed of 5/7 carbon rings and more unstructured than the previous case. It shows that once the solidification is stable, the atomic layout can't be changed easily under the effect of the adjacent layer potential, especially for layout with defects.

\subsubsection{Trilayer Structure}

\begin{figure}[ht]
 \begin{center}
\includegraphics[trim=10 10 10 10,width=3.3in]{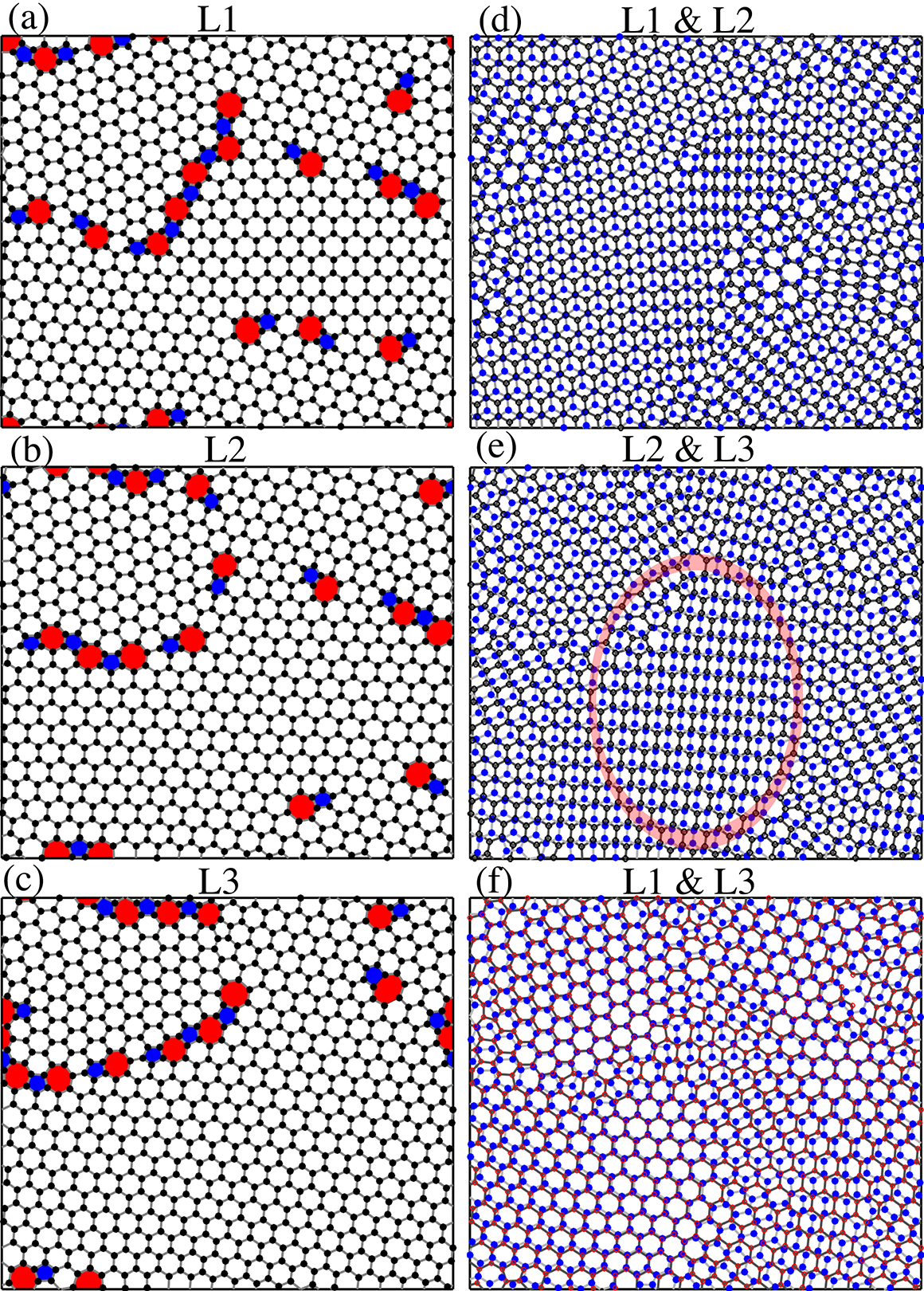}
\end{center}
\caption{(a) the layout of the atomic configuration for the bottom layer $L1$, (b)  layout of $L2$,  (c) layout of $L3$, (d)  $L1$ (blue dots) and $L2$  (black dots) stacking,  (e) $L2$ (black dots) and $L3$  (blue dots) stacking,  and (f) layouts of $L1$ (blue dots) and $L3$  (red dots).
\label{threeoneone}}
\end{figure}

\begin{figure}[ht]
 \begin{center}
\includegraphics[trim=10 10 10 10, width=3.3in]{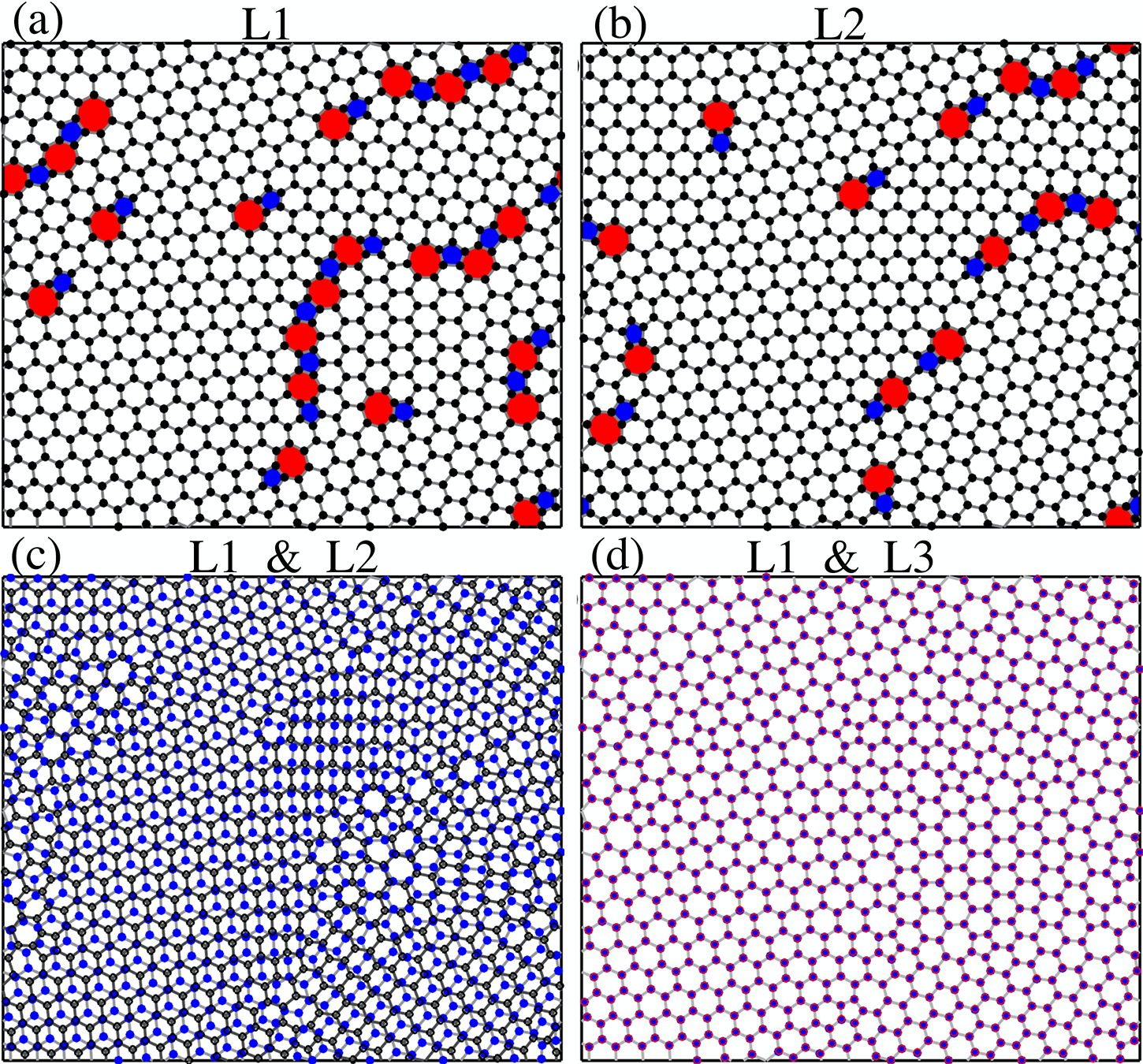}
\end{center}
\caption{
(a)  the layout of the atomic configuration for the bottom layer $L1$, (b) layout  $L2$, (c) layout of $L1$ (blue dots) and $L2$  (black dots), (d) layout of $L1$ (blue dots) and $L3$  (red dots). 
\label{threetwoone}
}
\end{figure}

Next we study the trilayer structures. Again we start with a well structured bottom layer ($L1$), then the second layer ($L2$) is added into the system with random initial condition, and finally the top layer ($L3$) after the solidification of $L2$. The results are very similar to the first case of bilayer structures, and $L3$ is almost exactly the same as $L1$. The $L1$ \& $L2$ stacking are perfect AB/BA stacking and so are the $L2$ \& $L3$ stacking.

\begin{figure}[ht]
 \begin{center}
\includegraphics[trim=10 10 10 10, width=3.3in]{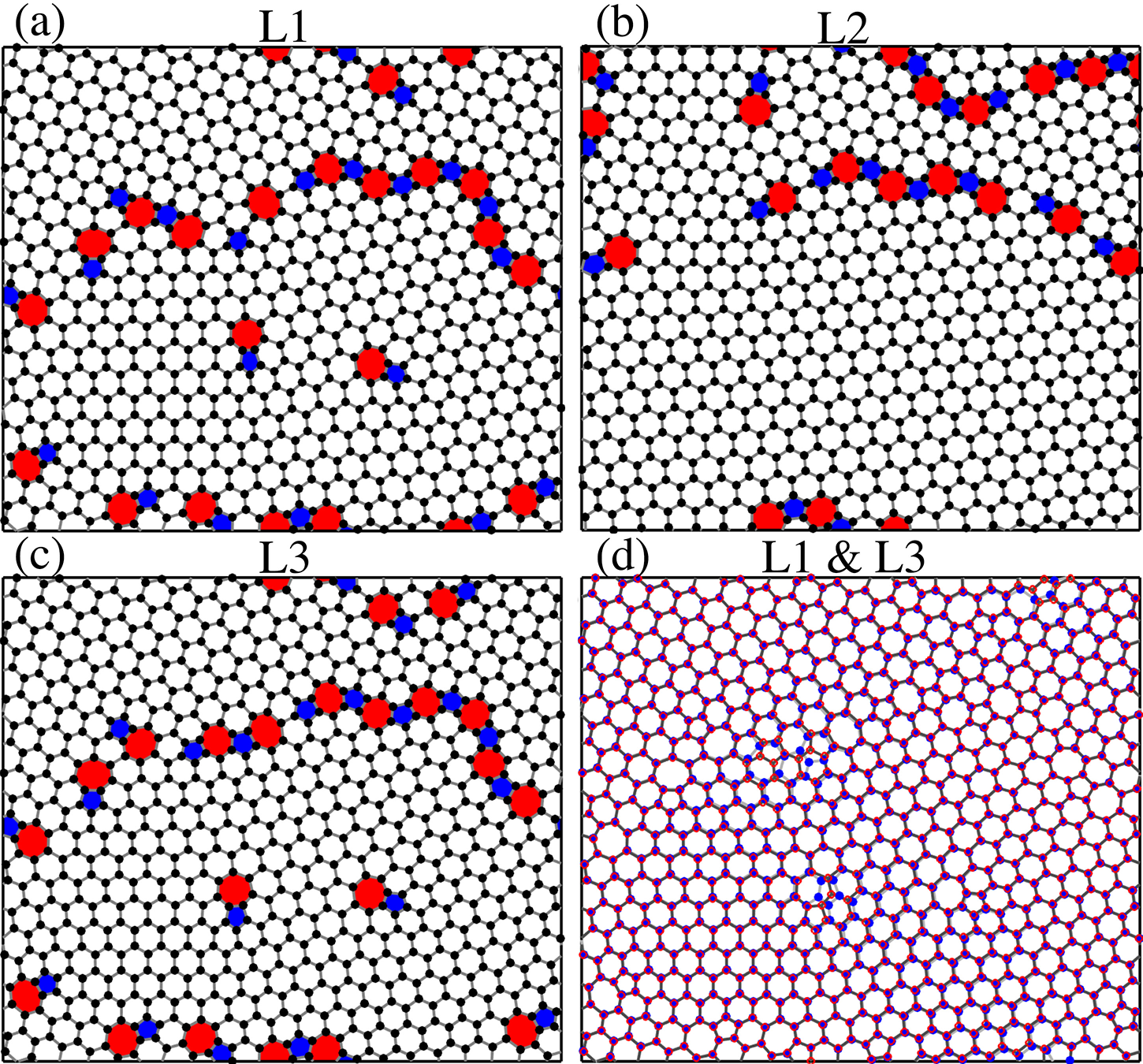}
\end{center}
\caption{(a) layout of $L1$, (b) layout of $L2$, (c)  layout of $L3$,  and (d) layouts of $L1$ (blue dots) and $L3$  (red dots). \label{threeslow}}
\end{figure}

\begin{figure}[ht]
 \begin{center}
\includegraphics[trim=10 10 10 10, width=3.3in]{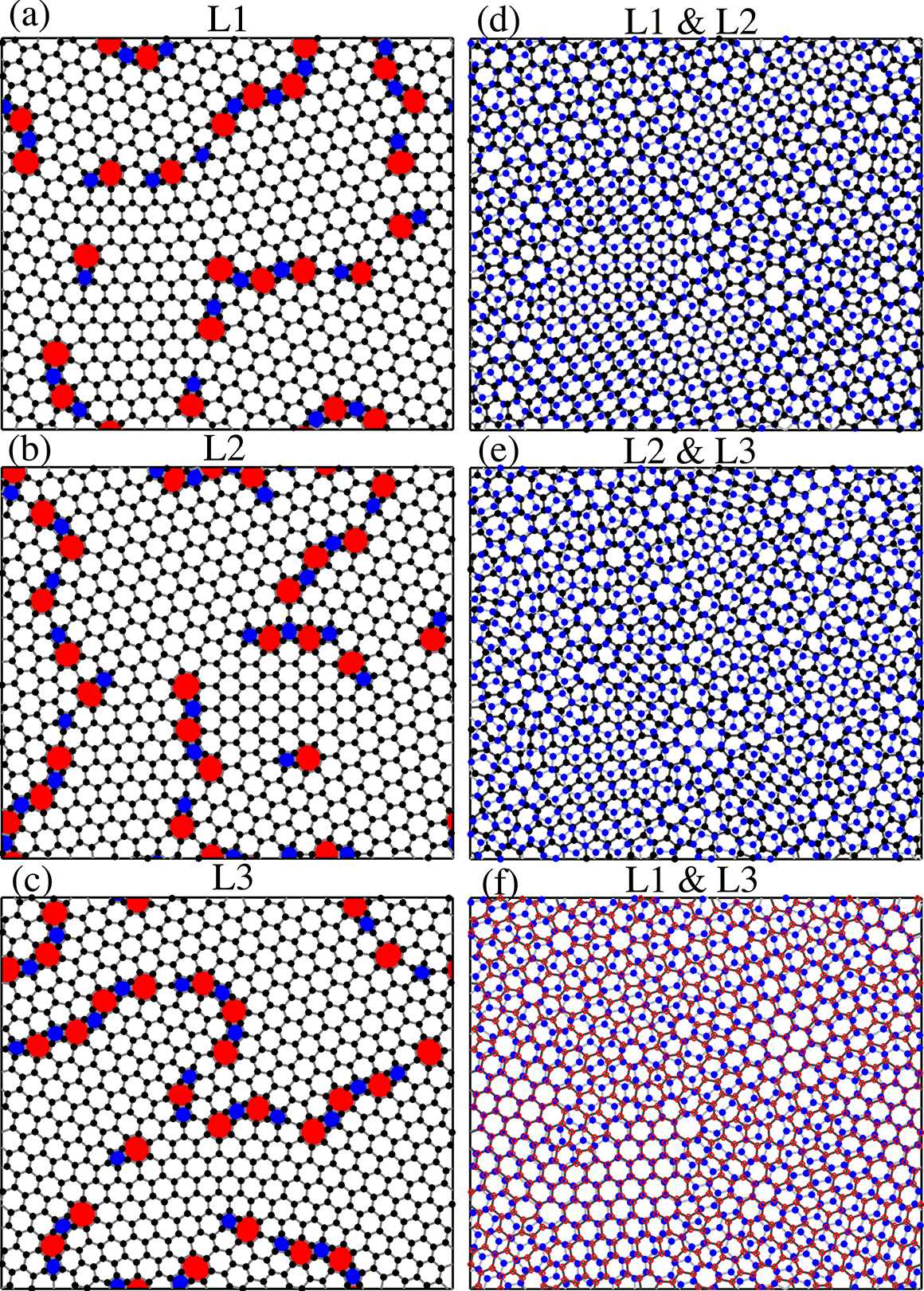}
\end{center}
\caption{(a) the layout of the atomic configuration for the bottom layer $L1$, (b)  layout of$L2$, (c) layout of $L3$,  (d) $L1$ (blue dots) and $L2$  (black dots) stacking, (e) $L2$ (black dots) and $L3$  (blue dots) stacking,  and (f) layouts of $L1$ (blue dots) and $L3$  (red dots).
\label{threesame}}
\end{figure}

The second case is that the initial condition of $L1$ is a constant (0.3) with small perturbations of Gaussian noise. After $L1$ is solidified, $L2$ is added into the system with similar initial condition, and finally the $L3$. The layouts of the atomic configuration for all the three layers are as shown in Fig. \ref{threeoneone}.  It turns out that the top layer is highly affected by the bottom layer, for both $L2$ \& $L1$ and $L3$ \& $L2$. However, $L3$ is away from exactly the same as $L1$, as shown in Fig. \ref{threeoneone} (f) that only about one third of the domain atoms on $L1$ (blue dots) and $L3$ (red circles) are close to overlapping in the $x-y$ plane. 

We further investigated into the details of the differences between the bilayer structure and the trilayer structure. As shown in Fig. \ref{threeoneone} (d) and (e). The $L1$ \& $L2$ stacking is similar to the bilayer case (in Fig \ref{twooneone} (d)), that more than half of domain are AB/BA stacking. However, the $L2$ \& $L3$ stacking are not the same, since nearly one fourth of the domain around the mid-bottom, i.e. the region circled by a red ellipse, is away from AB/BA stacking. Nevertheless, the atomic layouts for both $L2$ and $L3$ among this region are well structured, i.e. neatly arranged hexagons along the same direction. This could be explained by the following two reasons. Firstly, the inter-layer interactions dominate the process that $L3$ evolves to a well structured layout for the bottom half of the domain. Secondly, the AB/BA stacking between $L1$ and $L2$ prevents $L2$ layer further evolve to a more neatly arranged hexagon structure and AB/BA stacking with $L3$. Note that the stacking order of $L2$ \& $L3$ in the red ellipse is kind of a transition state between AB and BA stacking on the left and right.

The third case is that we start with a well structured $L1$ and then $L2$ and $L3$ are added into the system at the same time.  The final results are the same as case one, that both $L1$ \& $L2$ stacking and $L2$ \& $L3$ stacking are perfect AB-BA stacking.

The fourth case is similar to case 3, except that the initial condition of $L1$ is of constant value (0.3) with small perturbation of Gaussian noise. $L2$ and $L3$ are added into the system together after the solidification of $L1$.  The results are as shown in Fig. \ref{threetwoone}.  Unlike case two, here $L1$ and $L3$ are almost identical, as shown in Fig. \ref{threetwoone} (d). This is because at the beginning of the solidification, the phase field $L2$ mimic the opposite of $L1$, i.e. $\phi_2\sim -\phi_1$, and the phase field of $L3$ mimic the opposite of $L2$, which is  $L1$, i.e. $\phi_3\sim -\psi_2 \sim\phi_1$. So the solidification of $L3$ is highly affected by $L1$, which is not true for case two that $L3$ is not affected directly by $L1$. 

The fifth case is similar to case four, except that the effective mobility $M_\psi$ for $L3$ is smaller, e.g. $M_{\psi_3}=\frac{1}{3}M_{\psi_2},$ i.e. the phase field of the top layer $L3$ diffuses slower than $L2$. The results are as shown in Fig. \ref{threeslow}, there are slight differences between $L_1$ and $L_3$. In a certain sense, case five is more realistic, that once part of $L2$ is solidified, carbon rings on $L3$ may also be solidified onto that region. Certainly, solidification of carbon rings on $L3$ is slower than on $L2$.  

Finally we run a speciall case that solidifications of all the three layers start at the same time, each with randomly generated initial condition with different Gaussian noise. As shown in Fig. \ref{threesame}, only a small part of the domain for each pair are close to AB/BA stacking order, as shown in Fig. \ref{threesame} (d) and (e). And $L1$ \& $L3$ are far away from consistent. 

See more details of each case in the supplementary materials.

\subsubsection{Multiple Layer system\label{s4}}

 \begin{figure}[ht]
 \begin{center}
\includegraphics[trim=10 10 10 10, width=3.3in]{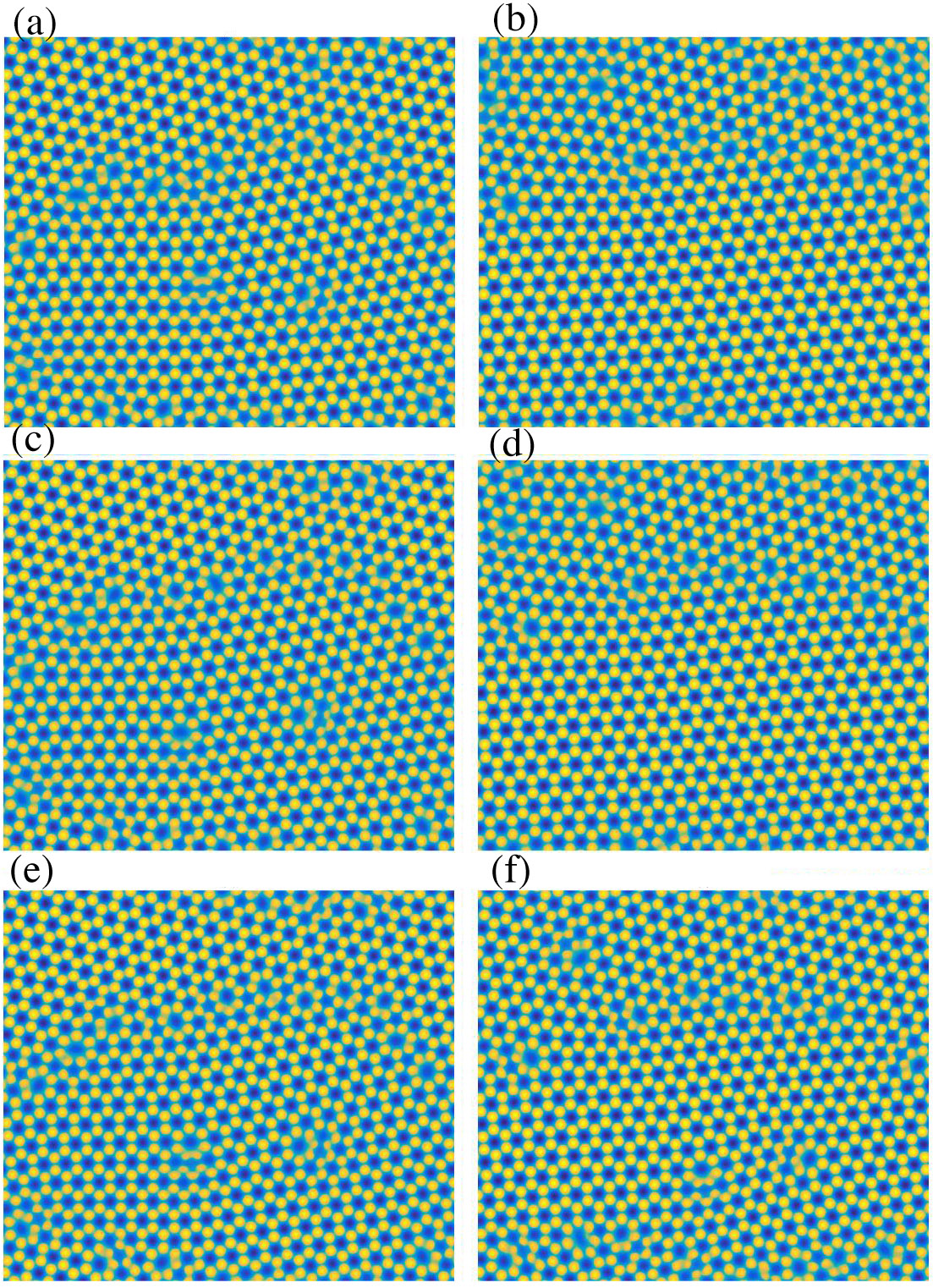}
\end{center}
\caption{(a) the layout of the phase field for the bottom layer $L1$, (b) layout of  $L2$, (c) layout  of  $L3$,
(d)  layout  of  $L4$, (e)  layout  of  $L5$, and (f) layout of $L6$.
 \label{six}}
\end{figure}

We also simulated multiple layer system, e.g. 4 layers, 5 layers, 6 layers, etc. Here we present a six-layer graphene structure, that one more layer is added into the system after the solidification of the previous layer. Note that by doing this, the 4 layer and 5 layer cases structures are also presented. The results are shown by Fig. \ref{six}. The odd layers $L1$, $L3$, and $L5$ follow one pattern and the even layers $L2$, $L4$, and $L6$ follow anther. As expected, there are small differences between the odd layers, e.g. the locations and shape of the $5/7$ carbon rings near the center are different for $L1$, $L3$, and $L5$. So do the even layers.

\section{Discussion}

In this paper, we build a new XPFC model for multilayer graphene by extending the XPFC method. The adjacent layer potential we use in this paper is similar to the generalized stacking fault energy, yet we use the corresponding phase filed instead. By doing this, the adjacent layer potential changes over time and systems of multiple layers can be easily built up.

We used the width of the AB-BA transition region of a long strip to determine the exact strength of adjacent layer potential, the results agree with atomistic simulations. We then simulate the multi-layer graphene structures. We test bilayers, trilayers, and multiple-layers. We've tried different initial conditions and various orders of construction. It turns out that once a base layer is formed, layers on the top will be affected and AB/BA stacking is more likely to be formed. Moreover, the defect grain boundaries will emerge around similar locations. For the well-structured bottom layer, the layers onto it are very likely to solidify into well-structured layouts.  By contrast, if there is not a well-structured base layer and multiple layers solidifies at the same time, adjacent layers will not be strongly correlated, and the grain boundary composed of 5/7 carbon rings are significantly different from each other. 

There is a natural next step to work with which is the rotational bilayer structural, for example the `1.1$^\circ$' magic degree rotation. Following the supercell method in \cite{trambly2010localization}, it is not hard to construct the initial phase field for all the layers. 

Although several set up of the simulations seem to be unphysical, we hope our numerical experiments help understanding the effect of the adjacent layer potential and the formation of multiple layer graphene structures. We believe that those `imaginary' cases are also valuable as those `real' ones, since they can't be easily testified by the experiments.


\begin{thebibliography}{plain} 

\bibitem{Oostinga2008naturematerials}J. B. Oostinga, H. B. Heersche, X. Liu, A. F. Morpurgo, and L. M. K. Vandersypen, ``Gate-induced insulating state in bilayer graphene devices,'' Nature Materials, vol. 7, p. 151-157, 2008.

\bibitem{butz2014dislocations} B. Butz, C. Dolle, F. Niekiel, K. Weber, D. Waldmann, H. B. Weber, B. Meyer, and E. Spiecker, ``Dislocations in bilayer graphene,'' Nature, vol. 505, no. 7484, p. 533, 2014.

\bibitem{ohta2006controlling}T. Ohta, A. Bostwick, T. Seyller, K. Horn, and E. Rotenberg, ``Controlling the electronic structure of bilayer graphene,'' Science, vol. 313, no. 5789, pp. 951-954, 2006.

\bibitem{yankowitz2019tuning} M. Yankowitz, S. Chen, H. Polshyn, Y. Zhang, K. Watanbe T. Taniguchi, D. Graf, A. F. Young, and C. R. Dean, ``Tuning superconductivity in twisted bilayer graphene,'' Science, vol. 363, no. 6431, pp. 1059-1064, 2019.\

\bibitem{cao2018unconventional} Y. Cao, V. Fatemi, S. Fang, K. Watanable, T. Taniguchi, E. Kaxiras, and P. Jarillo-Herrero, ``Unconventional superconductivity in magic-angle graphene superlattices,'' Nature, vol. 556, no. 7699, p. 43, 2018.

\bibitem{elder2004modeling} K. Elder and M. Grant, ``Modeling elastic and plastic deformations in nonequilibrium processing using phase field crystals,'' Physical Review E, vol. 70, no. 5, p. 051605, 2004.

\bibitem{elder2002modeling} K. Elder, M. Katakowski, M. Haataja, and M. Grant, ``Modeling elasticity in crystal growth,'' Physical review letters, vol. 88, no. 24, p. 245701, 2002.

\bibitem{meca2013epitaxial} E. Meca, J. Lowengrub, H. Kim, C. Mattevi, and V. B. Shenoy, ``Epitaxial graphene growth and shape dynamics on copper: phase-field modeling and experiments,'' Nano letters, vol. 13, no. 11, pp. 5692-5697, 2013.

\bibitem{Seymour2016Structural}  M. Seymour and N. Provatas, ``Structural phase field crystal approach for modeling graphene and other towdimensional structures,'' Physical Review B, vol. 93, no. 3, 2016.

\bibitem{Hirvonen2016} P. Hirvonen, M. M. Ervasti, Z. Fan, M. Jalalvand , M. Seymour, S. M. Vaez Allaei, N. Provatas, A. Harju, K. R. Elder, and T. Ala-Nissila, ``Multiscale modeling of polycrystalline graphene: A comparison of structure and
defect energies of realistic samples from phase field crystal models,'' Physical Review B, vol. 94, no. 3, p. 035414, Jul. 2016.

\bibitem{Kai} K. Liu, David, J. Lowengrub, and M. Austa, ``Phase Field Crystal Method for Bilayer Graphene'' Submitted, 2020.

\bibitem{Zhou2015Van}S. Zhou, J. Han, S. Dai, J. Sun, and D. J. Srolovitz, ``Van der waals bilayer energetics: Generalized stacking-fault energy of graphene, boron nitride, and graphene/boron nitride bilayers,'' Physical Review B, vol. 92, no. 15, 2015.

\bibitem{Alden2013Strain} J. S. Alden, A. W. Tsen, P. Y. Huang, H. Robert, B. Lola, P. Jiwoong, D. A. Muller, and P. L. Mceuen, ``Strain solitons and topological defects in bilayer graphene,'' Proceedings of the National Academy of Sciences of the United States of America, vol. 110, no. 28, pp. 11 256-11 260, 2013.

\bibitem{trambly2010localization} G. Trambly de Laissardiere, D. Mayou, and L. Magaud, ``Localization of dirac electrons in rotated graphene bilayers,'' Nano letters, vol. 10, no. 3, pp. 804-808, 2010.

\end{thebibliography}
\end{document}